\documentclass[prl,amsmath,amssymb,showpacs,superscriptaddress,floatfix]{revtex4}
\usepackage{graphicx}
\usepackage{bm}
\usepackage{lineno,hyperref}

\usepackage{epsf}
\usepackage{epstopdf}
\usepackage[english,russian]{babel}
\begin{document}
\title{Anomalous behavior of two-dimensional Hertzian sphere system}

\author{Eu. A. Gaiduk}
\affiliation{Vereshchagin Institute of High Pressure Physics, Russian Academy of Sciences,
Kaluzhskoe shosse, 14, Troitsk, Moscow, 108840 Russia}

\author{Yu. D. Fomin \footnote{Corresponding author: fomin314@mail.ru}}
\affiliation{Vereshchagin Institute of High Pressure Physics, Russian Academy of Sciences,
Kaluzhskoe shosse, 14, Troitsk, Moscow, 108840 Russia}
\affiliation{Moscow Institute of Physics and Technology (National Research University), 9 Institutskiy Lane, Dolgoprudny, Moscow region, 141701, Russia}

\author{E. N. Tsiok}
\affiliation{Vereshchagin Institute of High Pressure Physics, Russian Academy of Sciences,
Kaluzhskoe shosse, 14, Troitsk, Moscow, 108840 Russia}

\author{V. N. Ryzhov}
\affiliation{Vereshchagin Institute of High Pressure Physics, Russian Academy of Sciences,
Kaluzhskoe shosse, 14, Troitsk, Moscow, 108840 Russia}

\date{\today}

\begin{abstract}
The anomalous behavior of a two-dimensional system of Hertzian
spheres with exponent $\alpha = 7/2 $ has been studied using the
method of molecular dynamics. The phase diagram of this system is
the melting line of a triangular crystal with several maxima and
minima. Water-like density and diffusion anomalies have been found
in the reentrant melting regions. Noteworthy, a density anomaly
has been observed not only in the liquid and hexatic but also
solid phase. The calculations of the phonon spectra of
longitudinal and transverse modes have yielded negative dependence
of the frequency of transverse modes on density along all
directions in the regions with a density anomaly. This indicates
an association of the density anomaly with transverse oscillations
of the crystal lattice. The regions of density and diffusion
anomalies have been drawn on the phase diagram. It has been found
that the stability regions of anomalous diffusion extend to
temperatures well above maximum melting point $T = 0.0058$ of the
triangular crystal. From analysis of the translational order
parameter, which decreases with increasing density in the
reentrant melting regions, the presence of a structural anomaly in
the system has been assumed.
\end{abstract}


\maketitle

\section{Introduction}

It is known that two-dimensional (2D) systems can demonstrate a
number of properties that are very different from those of
three-dimensional (3D) systems. The most common example is melting
of 2D crystals. While in a 3D space melting always occurs through
a first-order phase transition, in 2D crystals there are known at
least three different melting scenarios. To date, for microscopic
description of 2D melting there are three recognized scenarios
\cite {ufn, 3d1}: (i) the theory of Berezinskii - Kosterlitz -
Thouless - Halperin - Nelson - Young (BKTHNY), according to which
melting occurs via two continuous transitions with an intermediary
hexatic phase with quasi-long-range orientational order and
short-range translational order \cite {berez, kt, halpnel1,
halpnel2, young}; (ii) melting in one transition of the first
order  \cite {chui, ourjetp}; (iii) the crystal-hexatic phase
transition takes place by means of a continuous transition of the
Berezinskii - Kosterlitz - Thouless (BKT) type, whereas the
hexatic phase-isotropic liquid transition - through a first-order
transition \cite {bernard, krauth, dikstra, kapfer, alice}. This
makes 2D systems extremely complex and interesting to study.

In addition, it is known that some systems may display complex
anomalous behavior. The most striking example of such systems is
water, in which dozens of different anomalous properties were
found, such as density anomaly, diffusion anomaly, and others. In
the case of water, many anomalies occur in the existence domain of
liquid. However, anomalous properties may take place in the
crystalline phase as well. For example, there are many crystals
known to demonstrate a density anomaly (in the case of crystals,
the term "a negative coefficient of thermal expansion" is commonly
used).

Many models have been proposed in literature to qualitatively
explain the anomalous behavior of water and other substances. An
important class of models leading to anomalous behavior in the
liquid phase is the class of systems with negative curvature
potentials (core-softened systems). Many different model
core-softened potentials have been proposed, some of which exhibit
anomalous behavior.

Another class of model systems in which anomalous behavior can
occur is associated with the so-called bounded potentials, i.e.,
those that do not show singularity at zero. Despite the seeming
simplicity of these potentials, they can demonstrate very complex
behavior, including reentrant melting, formation of cluster
crystals, numerous structural transitions, etc.

In a 2D space, systems with bounded potentials also show complex
behavior. For example, in the work \cite {gaussian} melting of a
system with the Gaussian potential was studied. Melting was shown
to occur through an intermediate hexatic phase. Besides, as in the
3D system with the Gaussian potential, anomalies of structure,
density and diffusion were found in 2D.

One of the most extensively studied systems with bounded potential
is the Hertzian sphere system (Hertzian spheres). It is defined by
the potential:

\begin{equation}\label{pot}
   U(r)=\varepsilon \left ( 1- r/ \sigma \right)^{\alpha}H(1-r),
\end{equation}
where $H(r)$ is the Heaviside step function and parameters
$\varepsilon$ and $\sigma$ set the energy and length scales. The
value of parameter $\alpha = 5/2$ corresponds to the Hertz problem
of energy during deformation of two elastic spheres.

The Hertzian sphere system with various values of $\alpha$ was
studied in both 3D and 2D. In the work \cite {c15}, the phase
diagram of a 3D system of Hertzian spheres with $\alpha = 5/2$ was
calculated. It was shown that a large number of structural
transitions among crystal phases of different symmetries occurred
in this system. At the same time, all phases showed a maximum on
the melting line. After that, in the work \cite {rosbreak} it was
shown that a diffusion anomaly and structural anomaly were also
observed in this system, moreover, there were two diffusion
anomaly areas in the system: at low and high densities. At the
same time, the diffusion anomaly at low density was more
pronounced. Later, qualitatively similar results were obtained for
a Hertzian sphere system with $\alpha = 2.0$ as well \cite
{lev1,lev2}.

The phase diagram of a 2D system of Hertzian spheres was studied
in even more detail. This issue was first addressed in the paper
\cite {miller}. In this article, the phase diagram of the system
was roughly estimated at three different values of the power of
$\alpha$: $\alpha = 3/2$, $5/2$ and $7/2$. Despite some
inaccuracies, this article detected the main trend of change in
the phase diagram with an increase in the power of $\alpha$: with
an increase in $\alpha$, the number of stable crystalline and
quasicrystalline phases in the system decreases. This result was
subsequently confirmed in the paper \cite {hertzqc}.

In the work \cite {hertzmelt}, the melting scenario of the
triangular phase with low density in a 2D system of Hertzian
spheres with $\alpha = 5/2$ was calculated. It was shown that the
melting line showed a maximum, in which the melting scenario
changed: at densities lower than the density at the melting line
maximum, a first-order transition from the hexatic phase to liquid
and a continuous transition from hexatic to crystal (the third
melting scenario) were observed, while at higher densities both
transitions were continuous according to the BKTHNY theory.
However, this paper examined the melting of only one triangular
phase at low densities. A complete phase diagram of the Hertzian
sphere system with $\alpha = 5/2$ was built in the article
\cite{molphys}. In the sequence of the density increase several
successive ordered phases are observed in the system, including
dodecagonal quasicrystal. The melting lines were evaluated for all
phases.

The system with $\alpha = 7/2$ deserves separate consideration. As
shown in \cite {miller}, there is only one crystal phase in this
system, a triangular crystal. At the same time, the melting line
looks complex. It is "wavy" in nature, that is, several maxima and
minima occur on the melting line. This result was significantly
expanded and revised in the article \cite {hz72}. In this paper it
was confirmed that only one crystal phase, a triangular crystal,
was observed in the system, and melting scenarios were calculated
for different parts of the melting line of this crystal. It was
shown that there were two melting scenarios in the system: the
third scenario, with a first-order transition from hexatic to
liquid, and the first scenario, with a continuous hexatic-liquid
transition according to the BKTHNY theory in all regions of
reentrant melting. In addition, a mechanism was proposed that
could lead to such non-standard behavior of the melting line.

Based on the complex behavior of the phase diagram of the 2D
Hertzian sphere system with $\alpha = 7/2$ and the presence of
regions with reentrant melting (a negative slope of the melting
line) of the triangular crystal, it can be assumed that certain
anomalous properties that were previously found in other 3D and 2D
systems with bounded potentials, will be observed in this system.
The purpose of this work is to search for such anomalies in the
studied system and to mark the areas of anomalous behavior on the
phase diagram.

\section{System and methods}

In our work, using the molecular dynamics method a 2D system of
Hertzian spheres with $\alpha = 7/2$ was simulated in a
rectangular box with periodic boundary conditions. The system
consisted of 20000 particles. A triangular crystal was selected as
the initial configuration. The simulation was carried out in the
canonical ensemble (constant number of particles N, volume V and
temperature T). The time step was $dt = 0.001$. First, 30 million
steps were performed to thermalize the system. Next, the system
was simulated for another 20 million steps to calculate its
properties. In conclusion, the system was simulated for more 20
million steps in the microcanonical ensemble (constant N, V and
internal energy E) to calculate the diffusion coefficient. The
diffusion coefficient was calculated from the mean square
displacement by the Einstein method.

The work also included calculation of the phonon spectra of the
crystal at different densities. The calculations were made in the
ground state using the Born-von Karman method.

\section{Results and discussion}

\begin{figure}

\includegraphics[width=8cm, height=8cm]{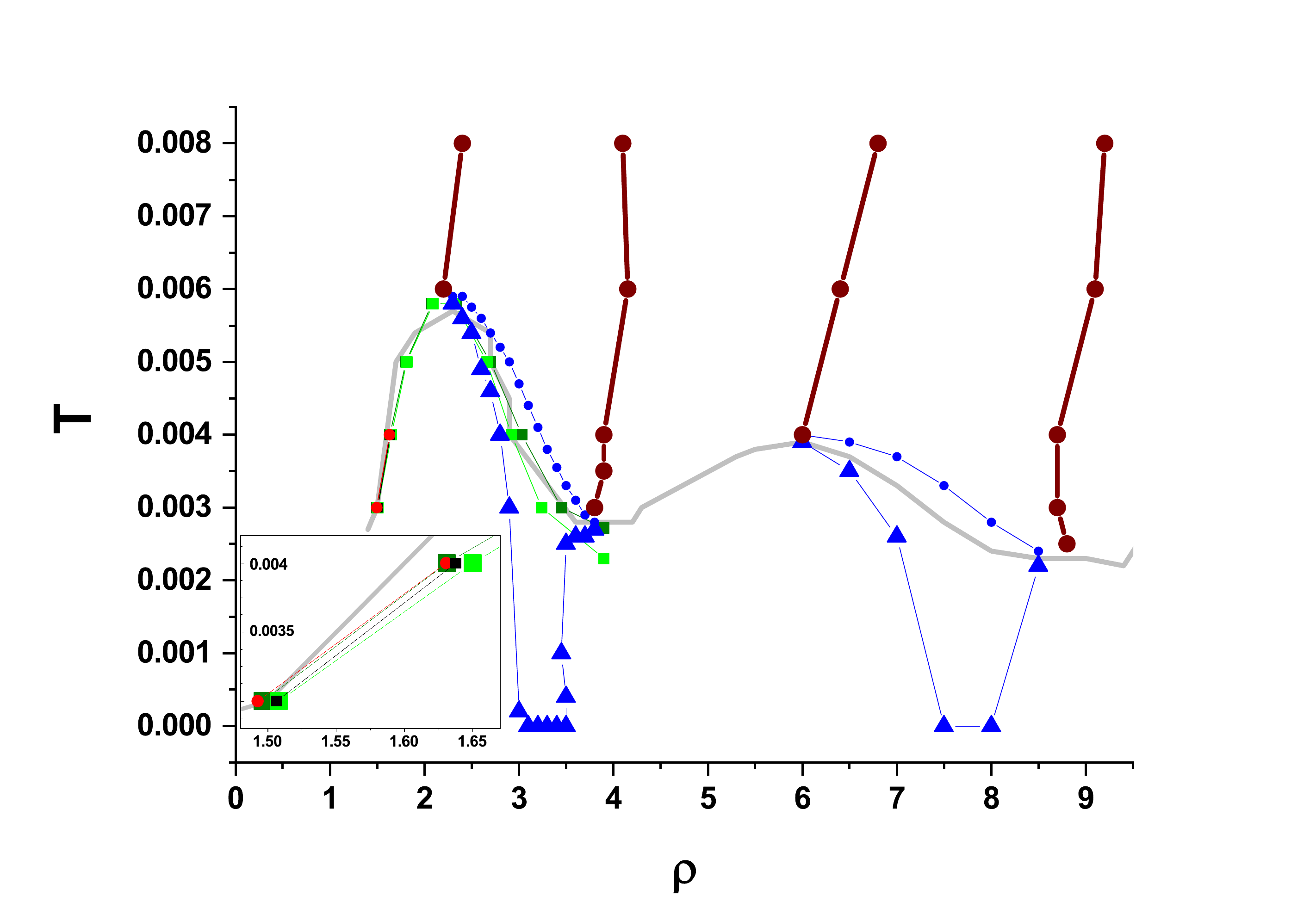}%

\caption{\label{pd} The phase diagram of the Hertzian spheres with
$\alpha = 7/2$ and the areas of density and diffusion anomalies.
The light grey line is the melting line from \cite {miller}. The
olive square is the stability limit of the hexatic phase. The
green square is the stability limit of the crystal. The red circle
is the beginning of the Mayer-Wood loop corresponding to a
first-order liquid-hexatic transition. The black square is the end
of the Mayer-Wood loop. The inset zooms in the region with the
third scenario (with a first-order transition from hexatic to
liquid). The blue triangle is the boundary of the density anomaly
area determined from the maximum on the isochores. The blue circle
is the boundary of the density anomaly area determined from the
minimum on the isochores. The wine circle is the boundary of the
diffusion anomaly areas.}
\end{figure}

The phase diagram of the system under consideration was calculated
in our previous work \cite {hz72}. Recall that this phase diagram
has a very unusual shape: despite the fact that there is only one
crystalline phase, the melting line has a wavy shape with several
maxima and minima. Part of the phase diagram in the area of
studied densities is shown in Fig. \ref {pd}. In the present work,
we continue studying the 2D Hertzian sphere system with $\alpha =
7/2$ and investigate into the origination of various anomalies in
it, which are observed in many systems with anomalous phase
diagrams.

\begin{figure}

\includegraphics[width=8cm]{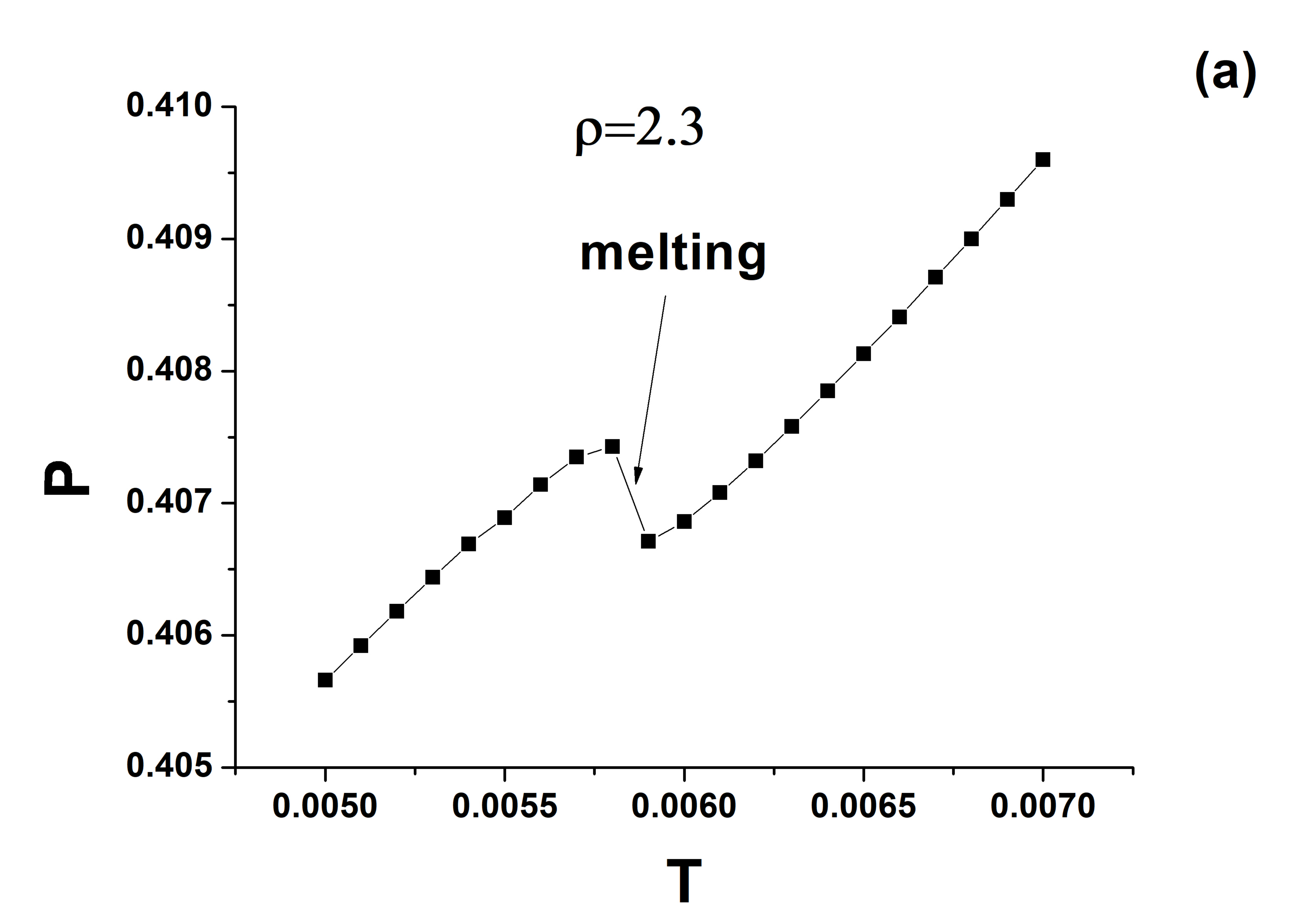}%
\includegraphics[width=8cm]{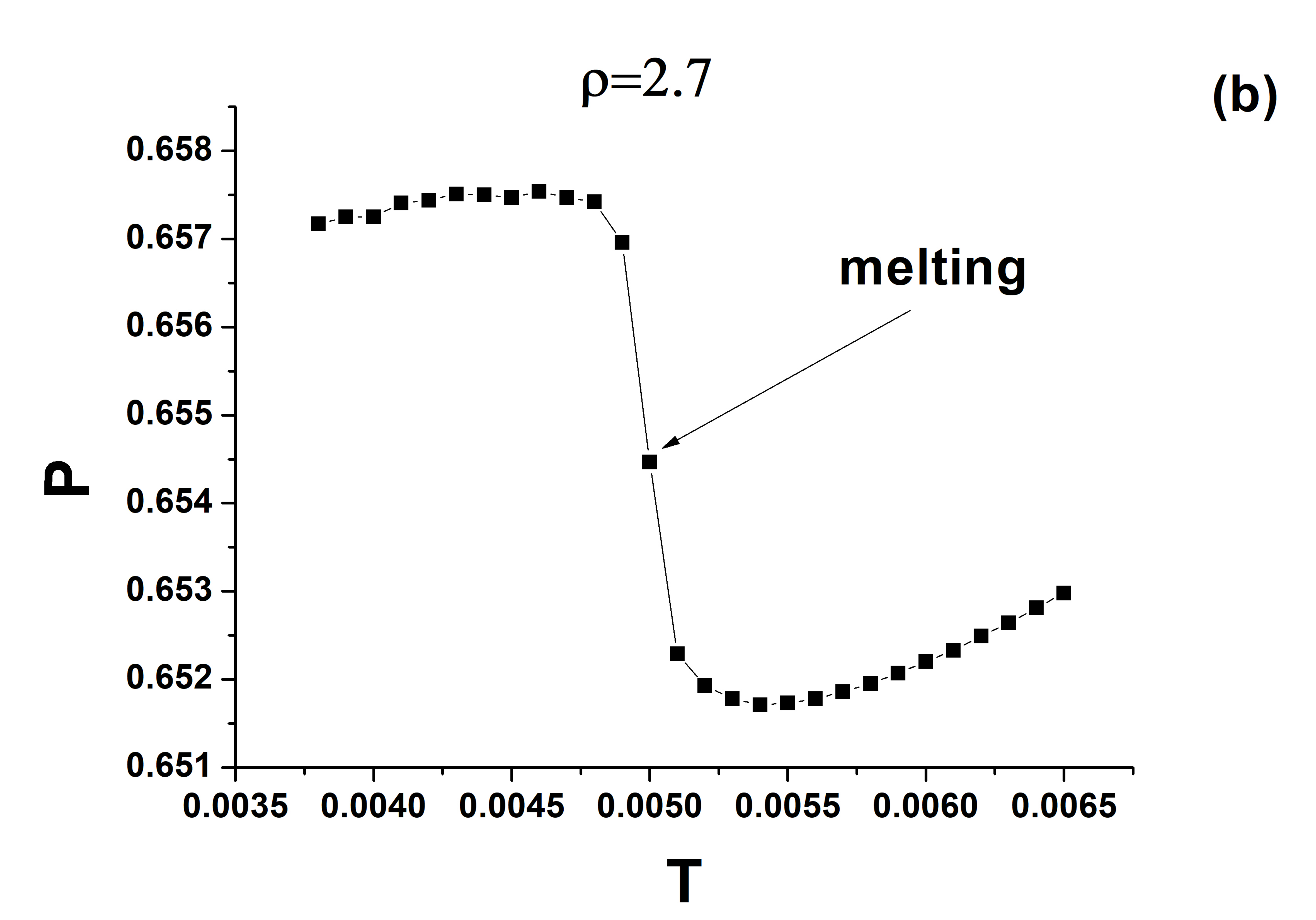}%

\includegraphics[width=8cm]{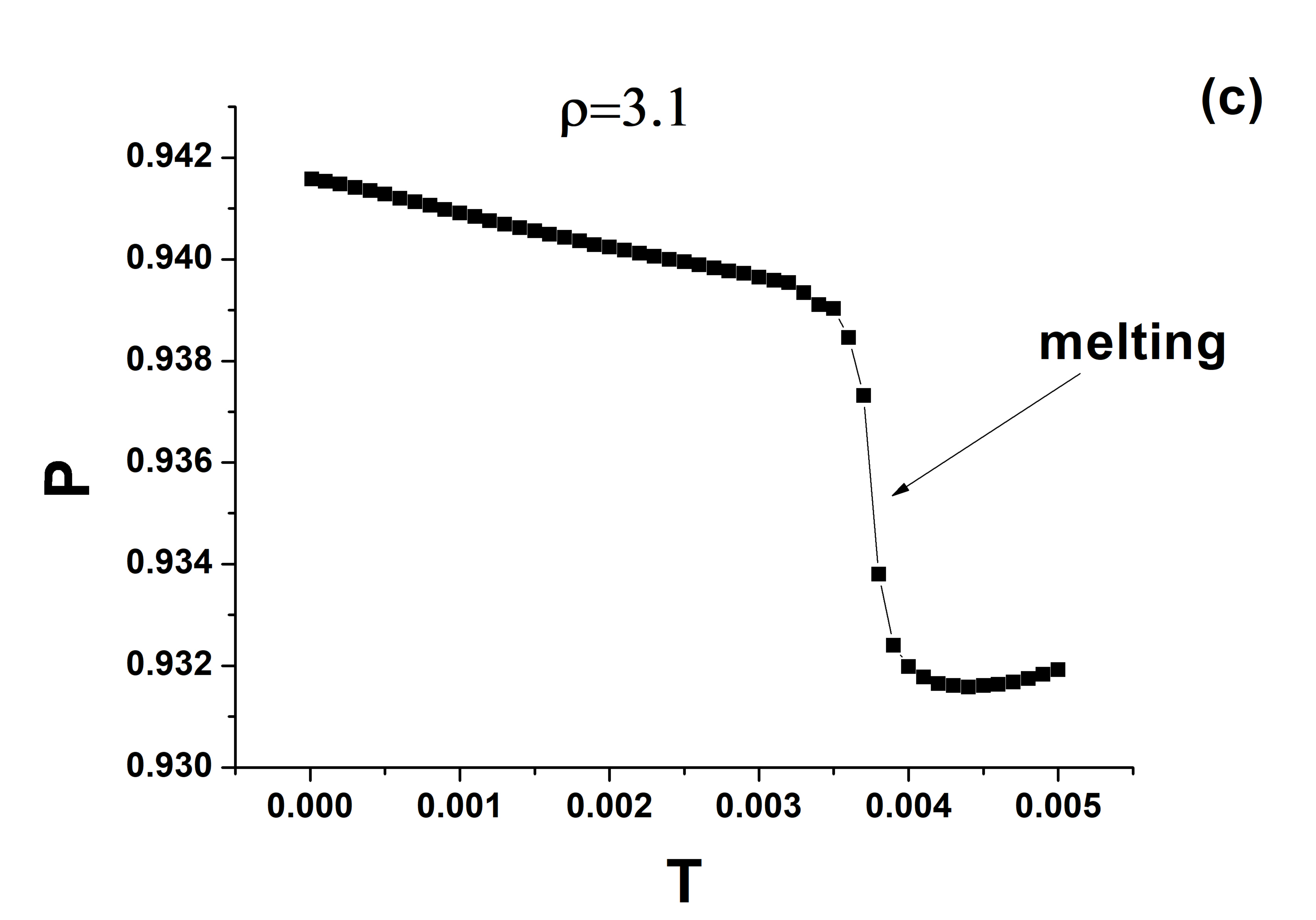}%
\includegraphics[width=8cm]{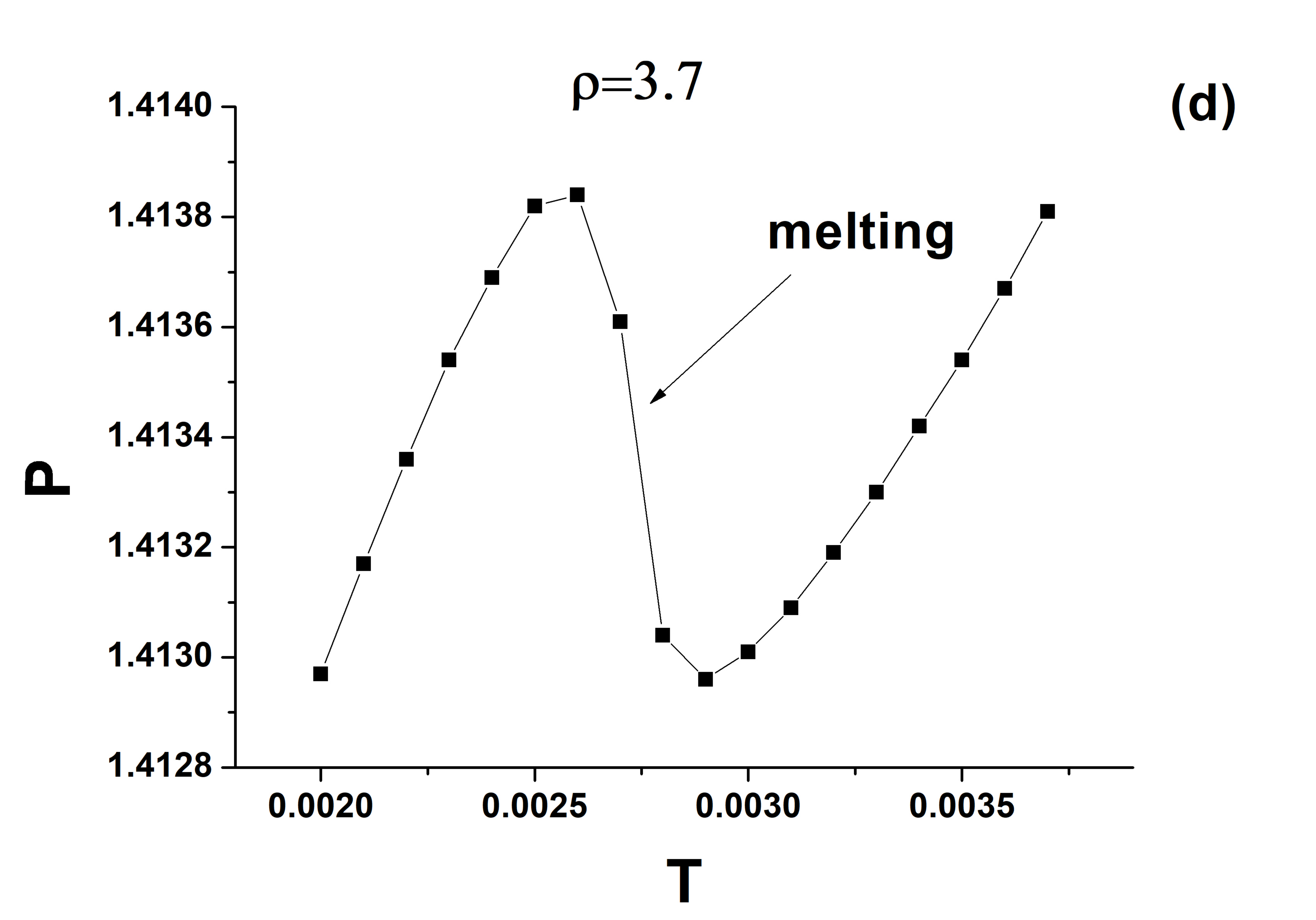}%

\includegraphics[width=8cm]{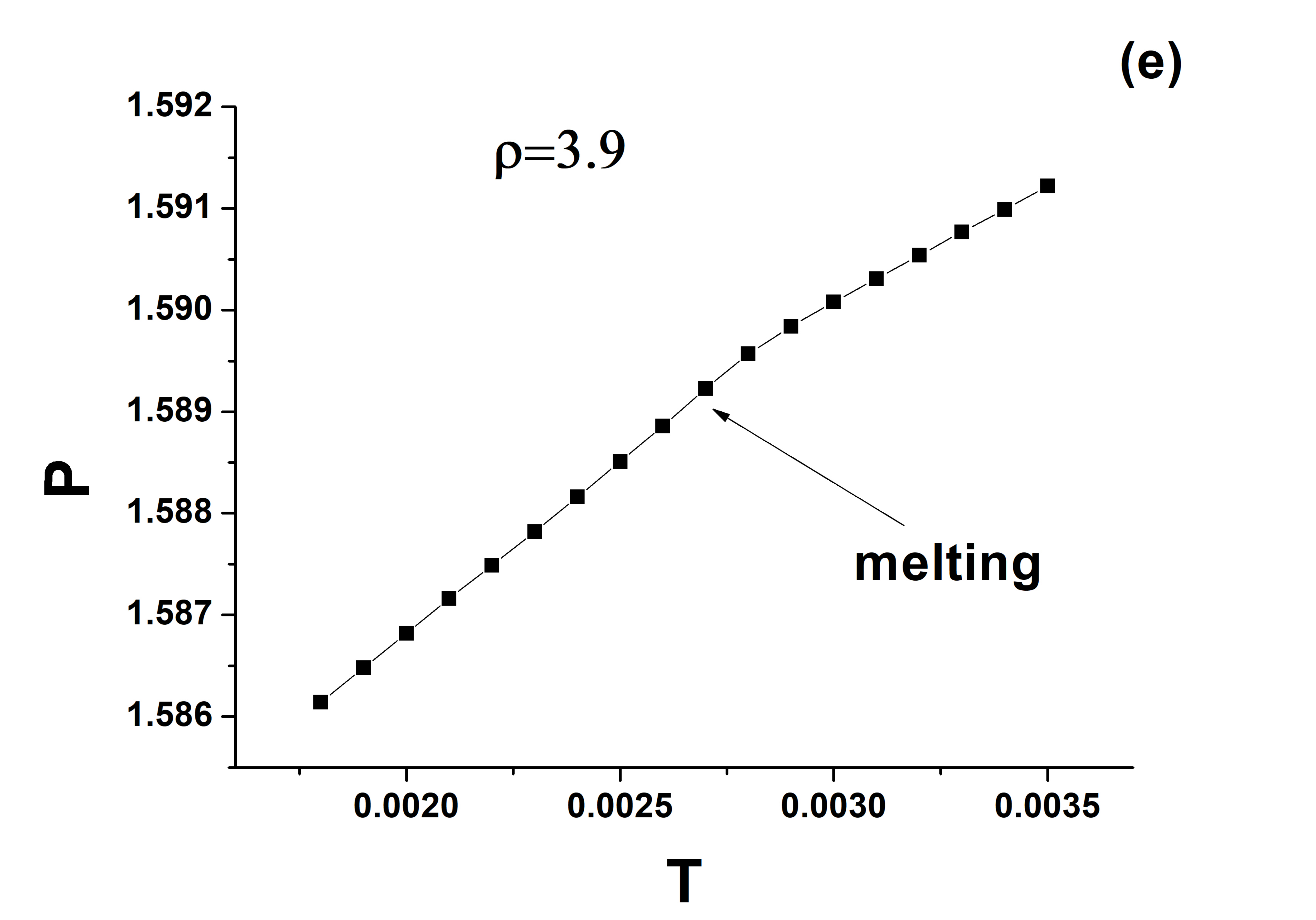}%

\caption{\label{isorho} The equations of state along several
isochores in the vicinity of reentrant melting of the first bump
on the phase diagram.}
\end{figure}

Let us begin the description of the obtained results with looking
for a density anomaly in the system whose physical meaning
consists in the negative coefficient of thermal expansion. Using
the thermodynamic relationship $\left (\frac {\partial P}
{\partial T }\right) _ V $ = $\alpha _ P $/$K _ T $, where
$\alpha _ P$ is the coefficient of thermal expansion and $K _ T$
is isothermal compressibility, and considering that $K _ T $ is
always positive and finite for systems in equilibrium outside a
critical point, it can be concluded that a density anomaly
corresponds to a decrease in pressure with an increase in
temperature along the isochores, and a minimum on this dependence
- to the boundary of the anomalous area [see Eq. (6) in
\cite{s135e}, \cite{anomwe}]. The derivative $\left (\frac
{\partial P} {\partial T }\right) _ V$ in the anomalous area
should be negative. Therefore, we studied the equations of state
along the isochores. Recall that in the areas of reentrant melting
under consideration, the crystal melts via two continuous BKT
transitions through the hexatic phase. Figures \ref {isorho} (a) -
(e) show a number of the system isochores. It can be seen that at
density $\rho = 2.3$, the pressure increases with increasing
temperature in both solid and liquid phases. Only in a very narrow
temperature range with width 0.0001 the pressure decreases.
Density $\rho = 2.3$ is near the maximum of the melting line,
where melting passes through a very narrow region of the hexatic
phase. Therefore, it can be assumed that the observed density
anomaly occurs in the orientationally ordered liquid-hexatic
phase. At density $\rho = 2.7$, the pressure decreases to a
minimum in the temperature region, the width of which is by an
order of magnitude greater than at $\rho = 2.3$. At the same time,
the density anomaly area includes part of the crystalline phase,
full hexatic and part of the liquid phase. On the equation of
state along isochore $3.1$, starting with temperature close to
zero, a monotonous pressure decrease is observed with an increase
in temperature, which corresponds to the anomaly of density in the
crystal. Then a region of an abrupt pressure drop appears as a
function of temperature, which mainly falls within the existence
domain of the hexatic phase. Next, a minimum appears on the
equation of state that corresponds to the boundary of the density
anomaly area in the liquid. With a further increase in
temperature, pressure increases, which indicates the disappearance
of the density anomaly in the liquid. Thus, at density $\rho =
3.1$ the domains of existence of the crystalline and hexatic
phases completely and of the liquid phase partially fall within
the area of the density anomaly. On isochore $\rho = 3.7$ the
maximum and minimum correspond to the density anomaly area, which
completely covers the existence domain of the hexatic phase and a
very small area of the liquid phase. It should be noted that such
a course of the equation of state along the above-mentioned
isochores, namely, a pressure drop with an increase in
temperature, is mainly characteristic of the reentrant melting
regions on the phase diagram. With a further increase in density
($\rho = 3.9$), normal behavior of pressure dependence on
temperature is observed, i.e., the density anomaly disappears.
Thus, we see that the density anomaly in the system under
investigation is observed not only in the liquid and hexatic but
also in solid phases.

\begin{figure}

\includegraphics[width=8cm]{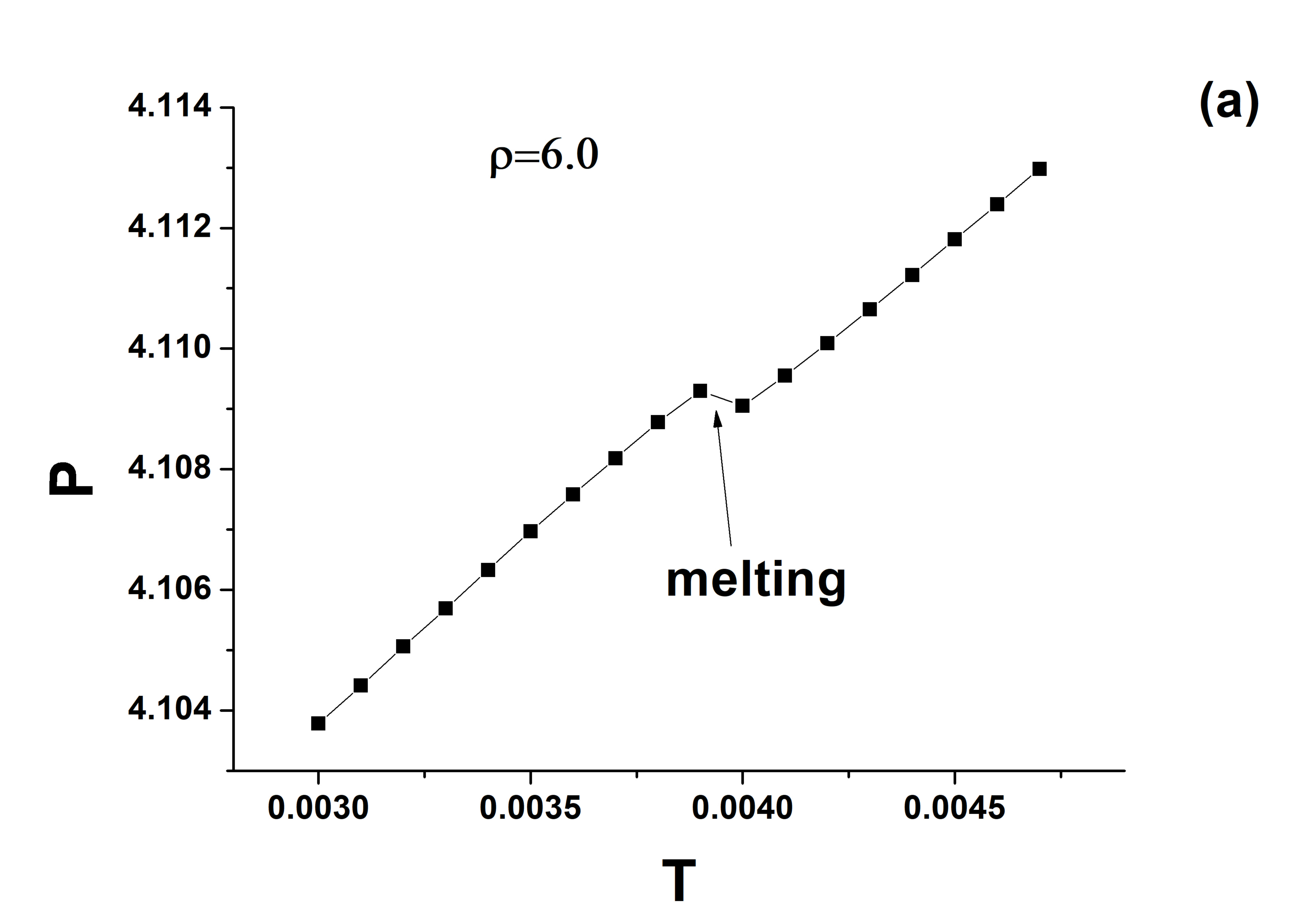}%
\includegraphics[width=8cm]{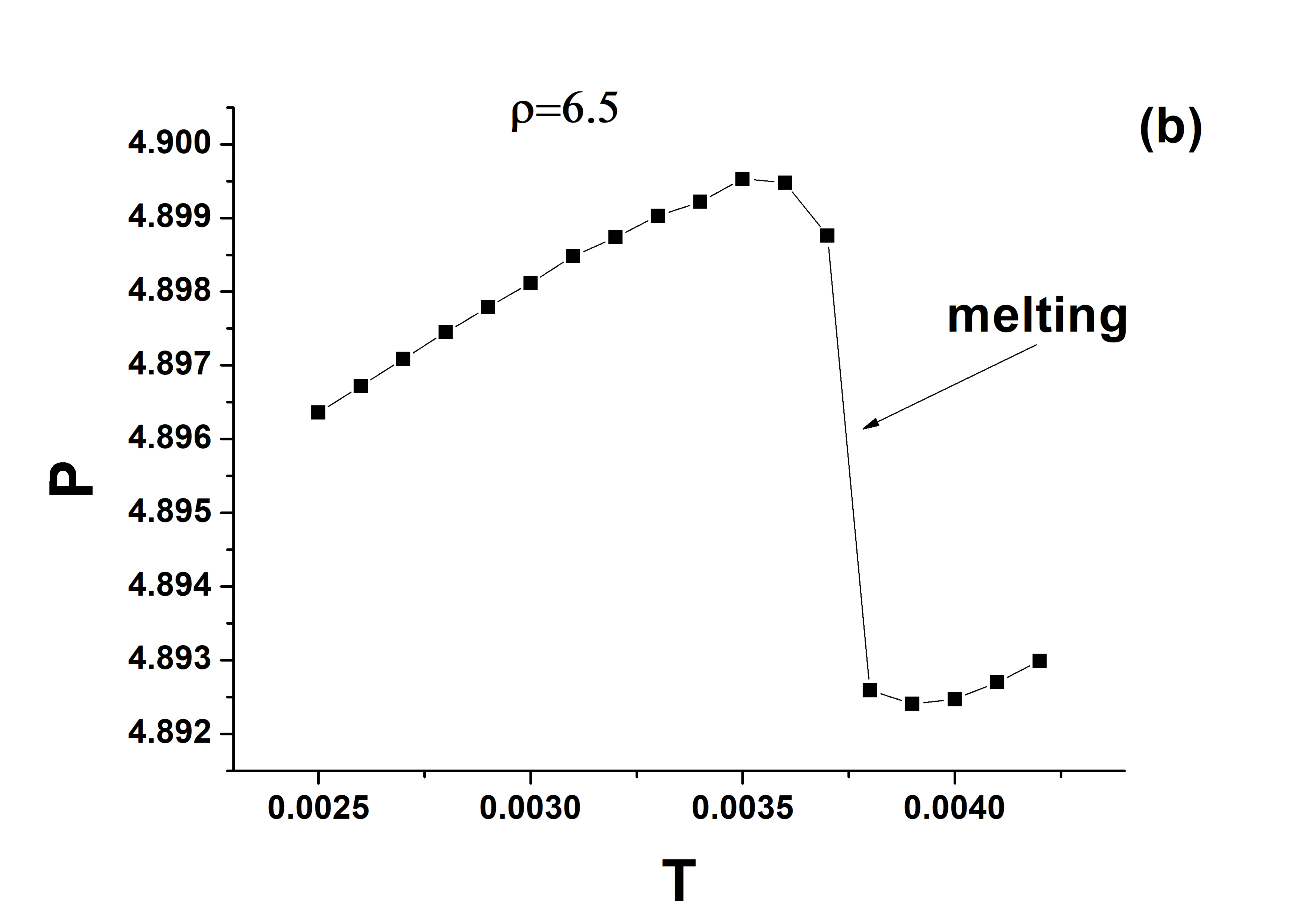}%

\includegraphics[width=8cm]{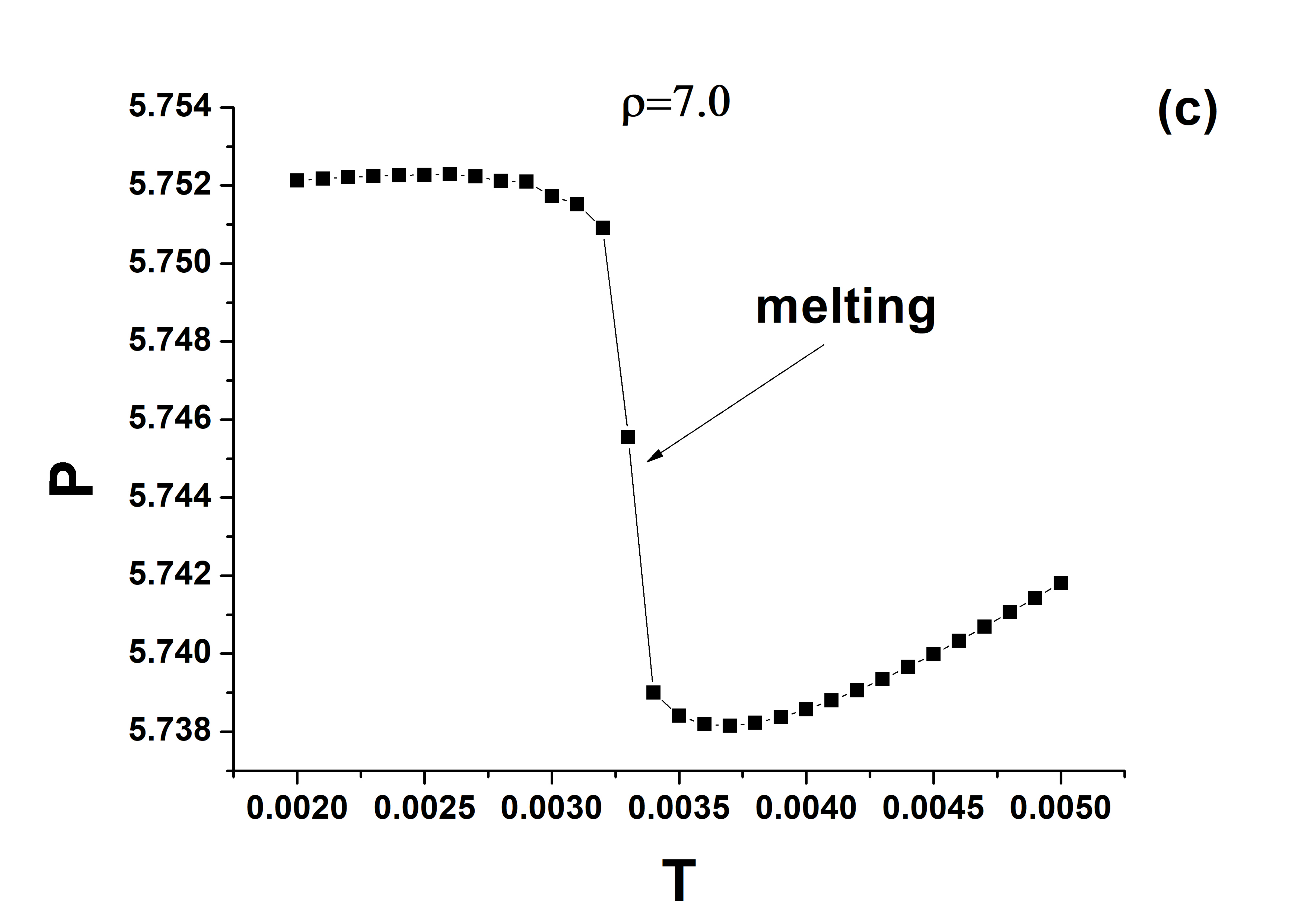}%
\includegraphics[width=8cm]{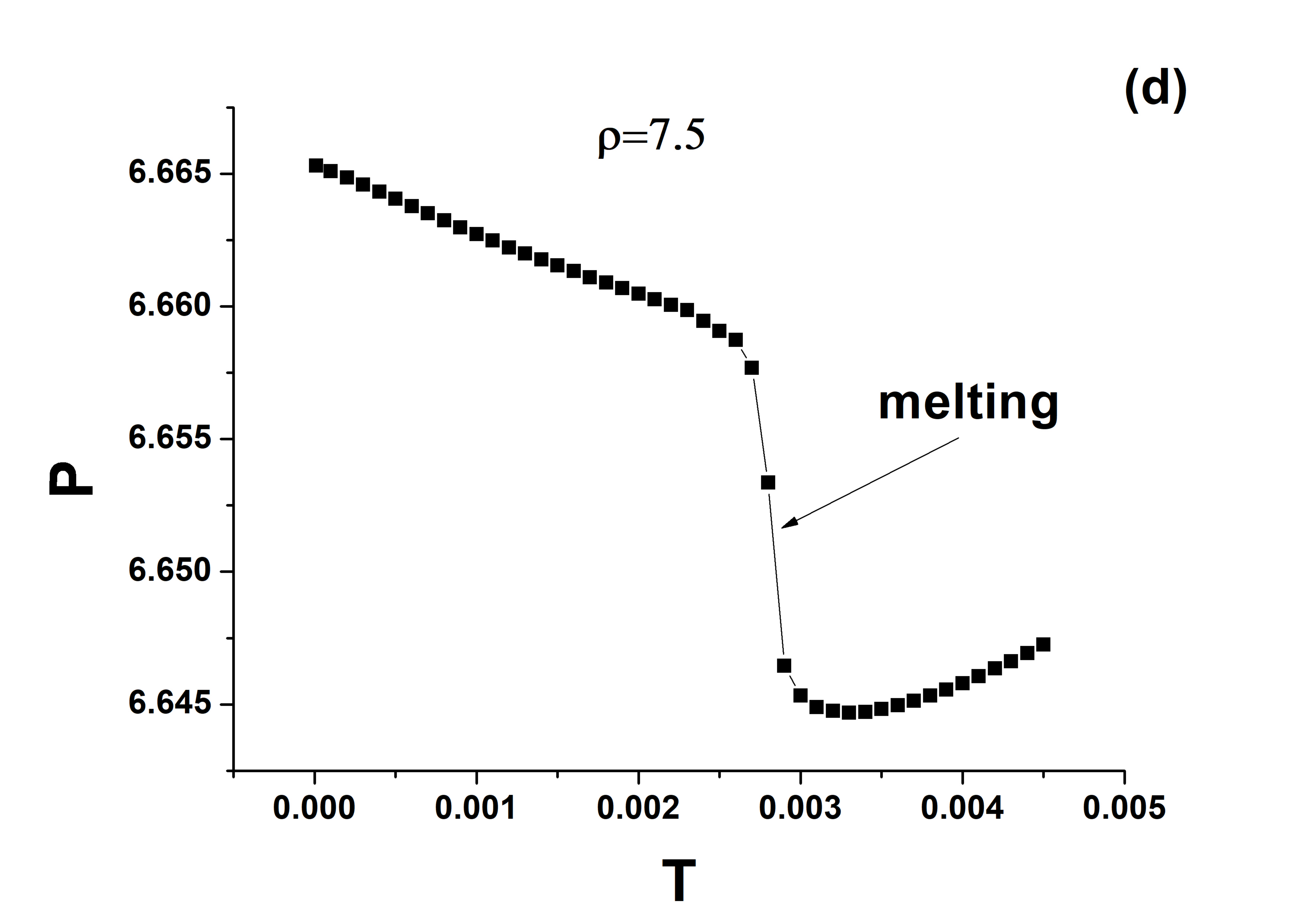}%

\includegraphics[width=8cm]{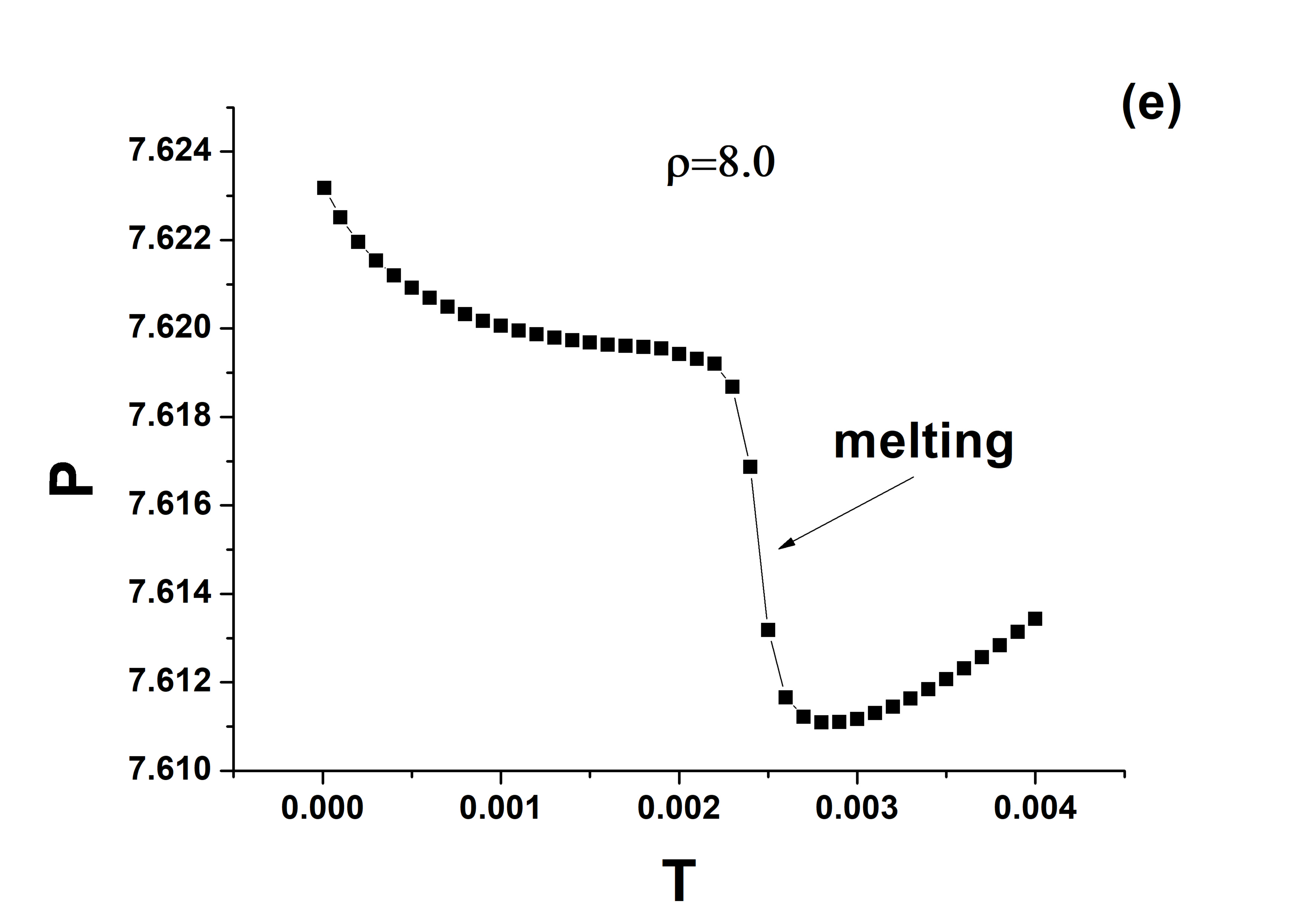}%
\includegraphics[width=8cm]{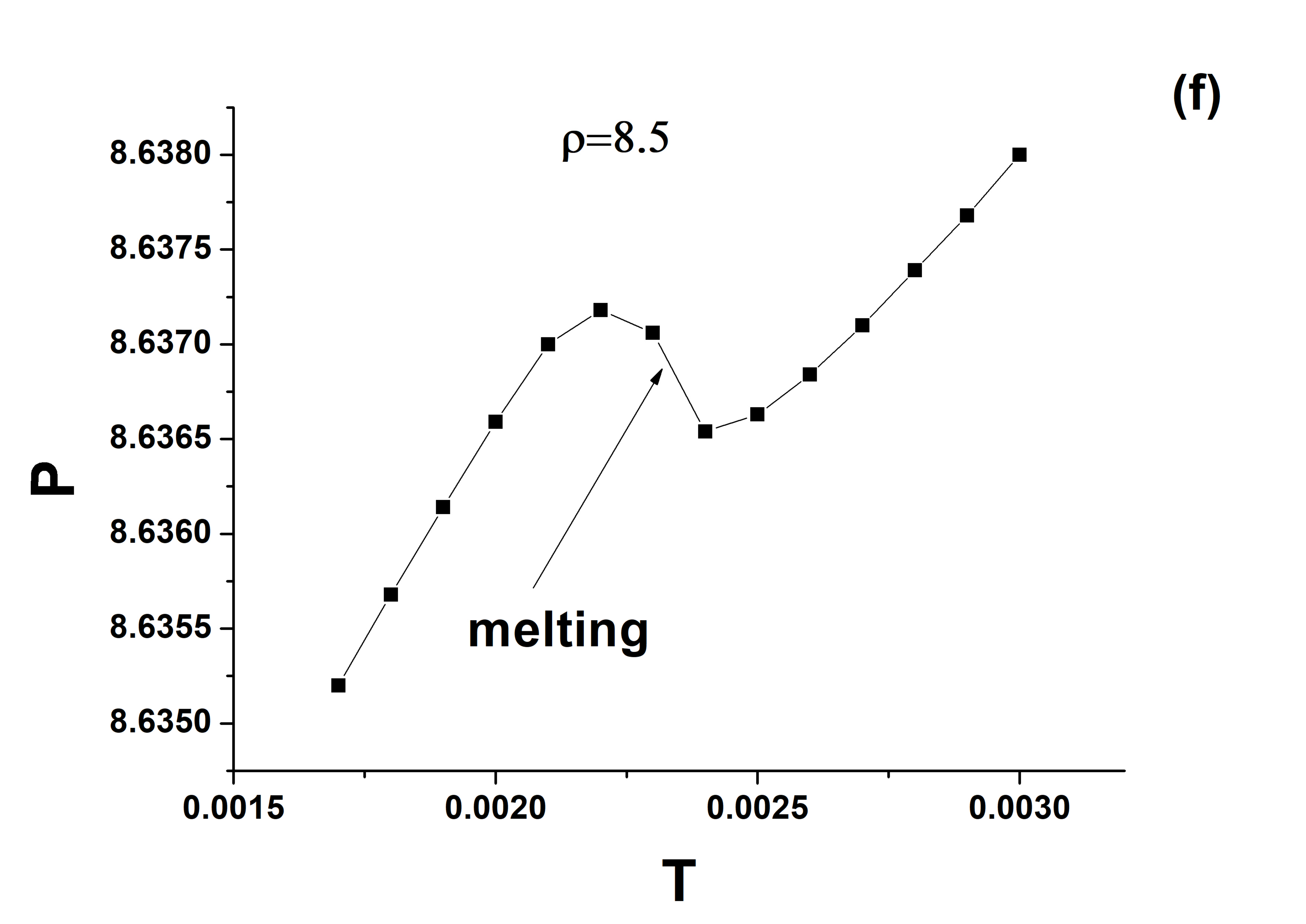}%

\caption{\label{isorho2} The equation of state along several
isochors in the vicinity of reentrant melting of the second bump
on the phase diagram.}
\end{figure}

Recall that the melting line of the system has a wavy shape. This
suggests that the behavior of anomalies may also be periodic.
Therefore, we studied the isochores corresponding to the right
branch of the melting line's second maximum. Corresponding graphs
are shown in Fig. \ref {isorho2}. We see that isochore behavior is
indeed similar to that at the densities around the first peak. A
density anomaly can appear in both solid and liquid phase
(densities $\rho = 7.5$ and $\rho = 8.0$). With a further increase
in density, the anomaly disappears.

\begin{figure}

\includegraphics[width=8cm]{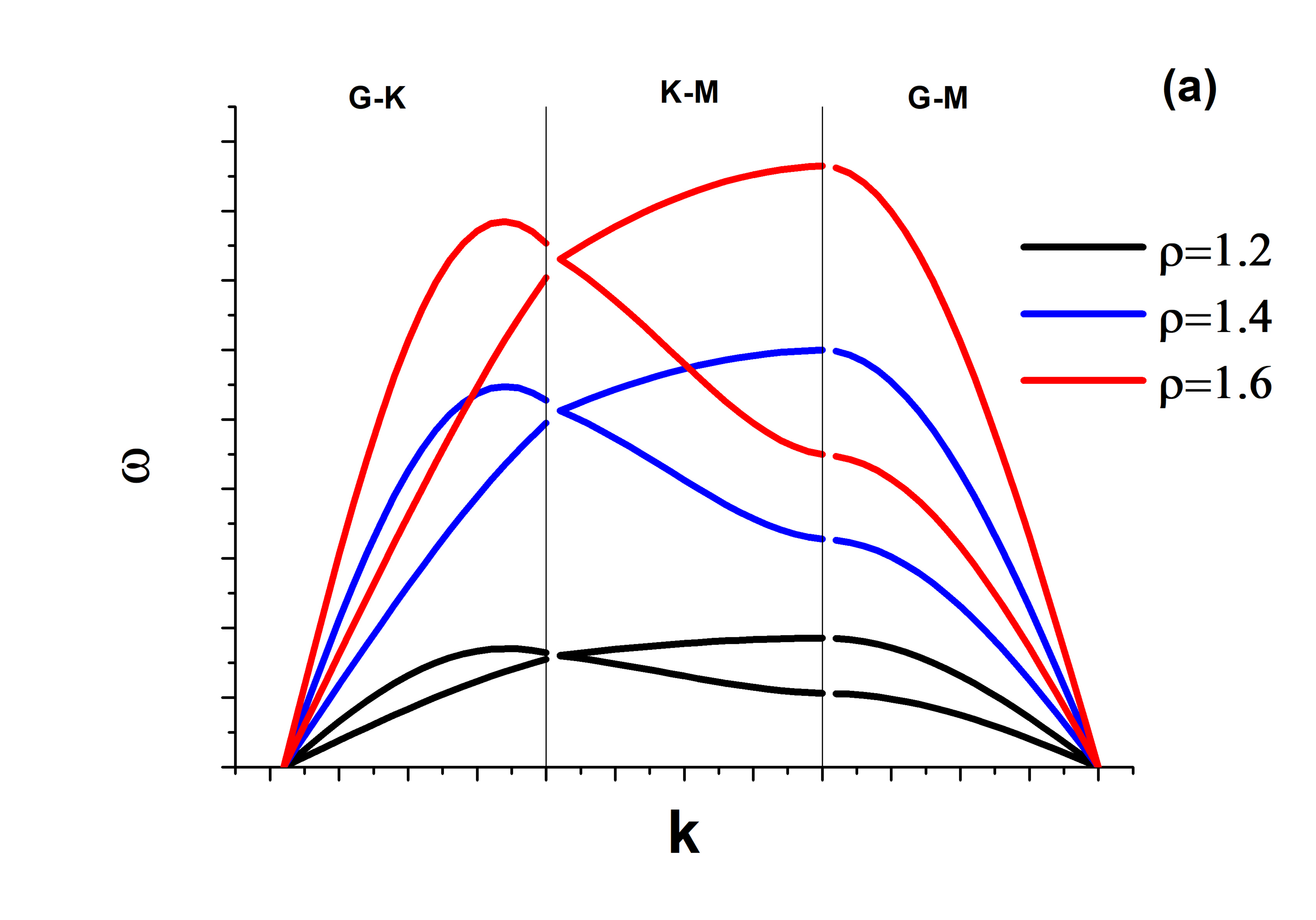}%
\includegraphics[width=8cm]{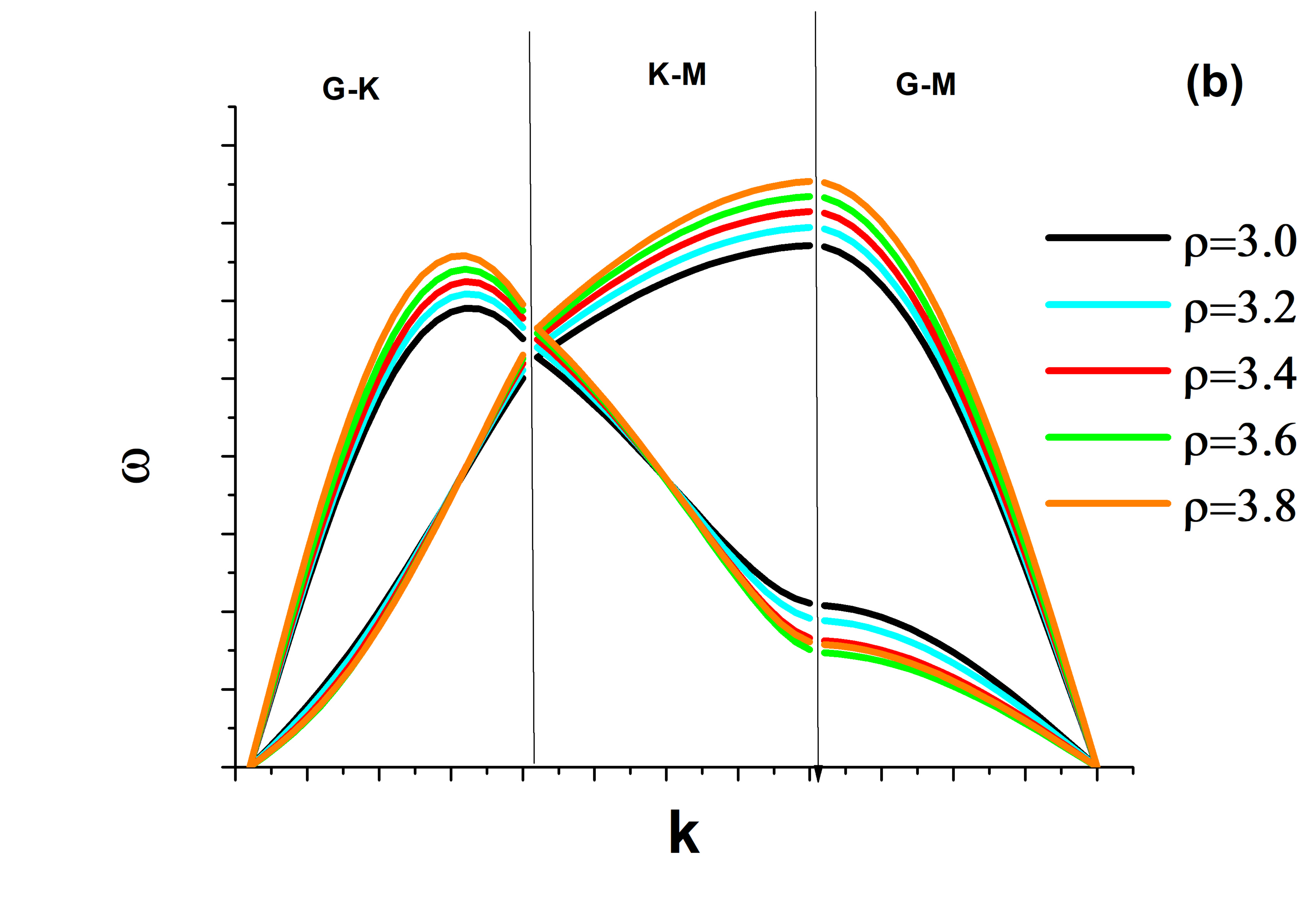}%

\includegraphics[width=8cm]{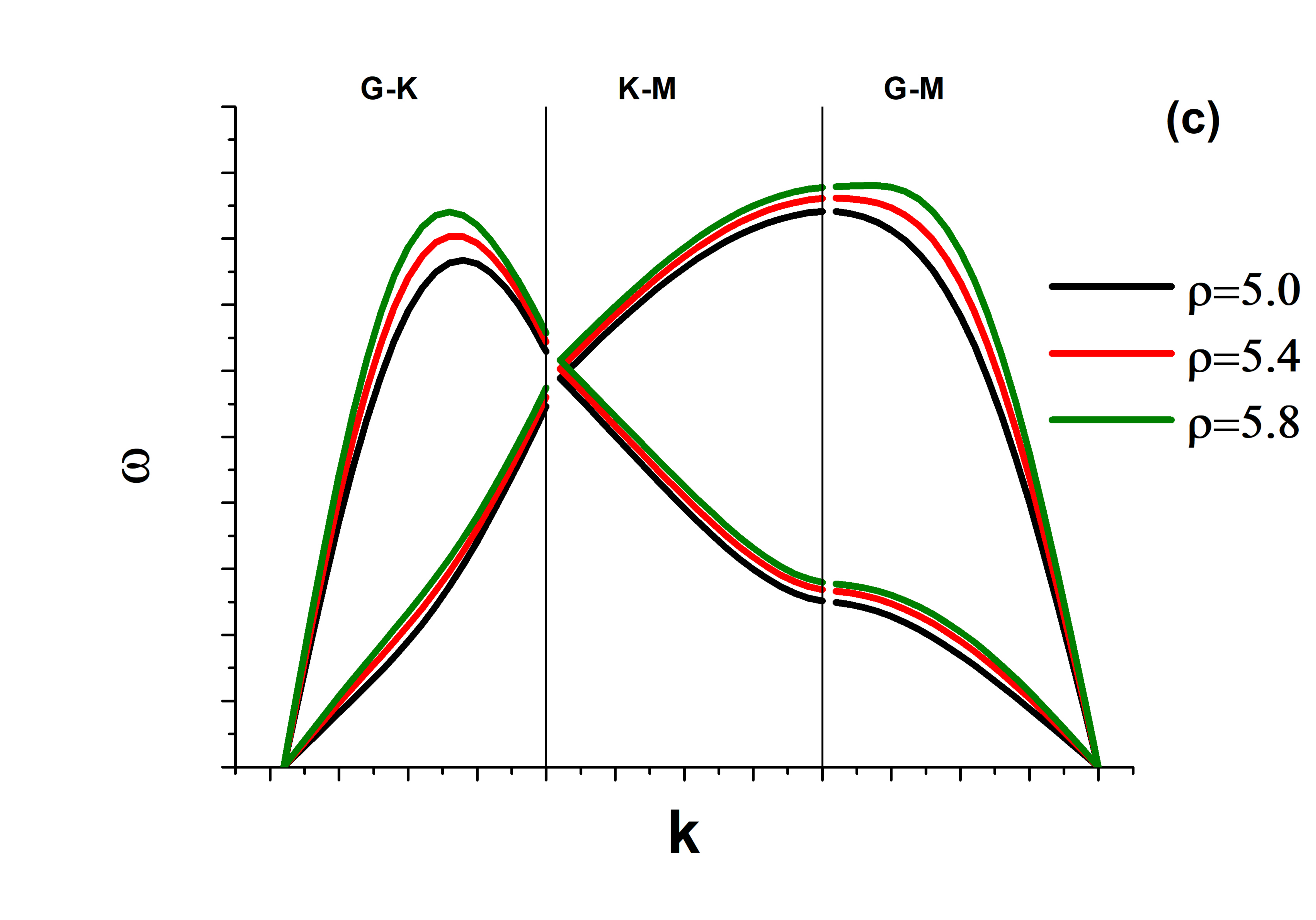}%
\includegraphics[width=8cm]{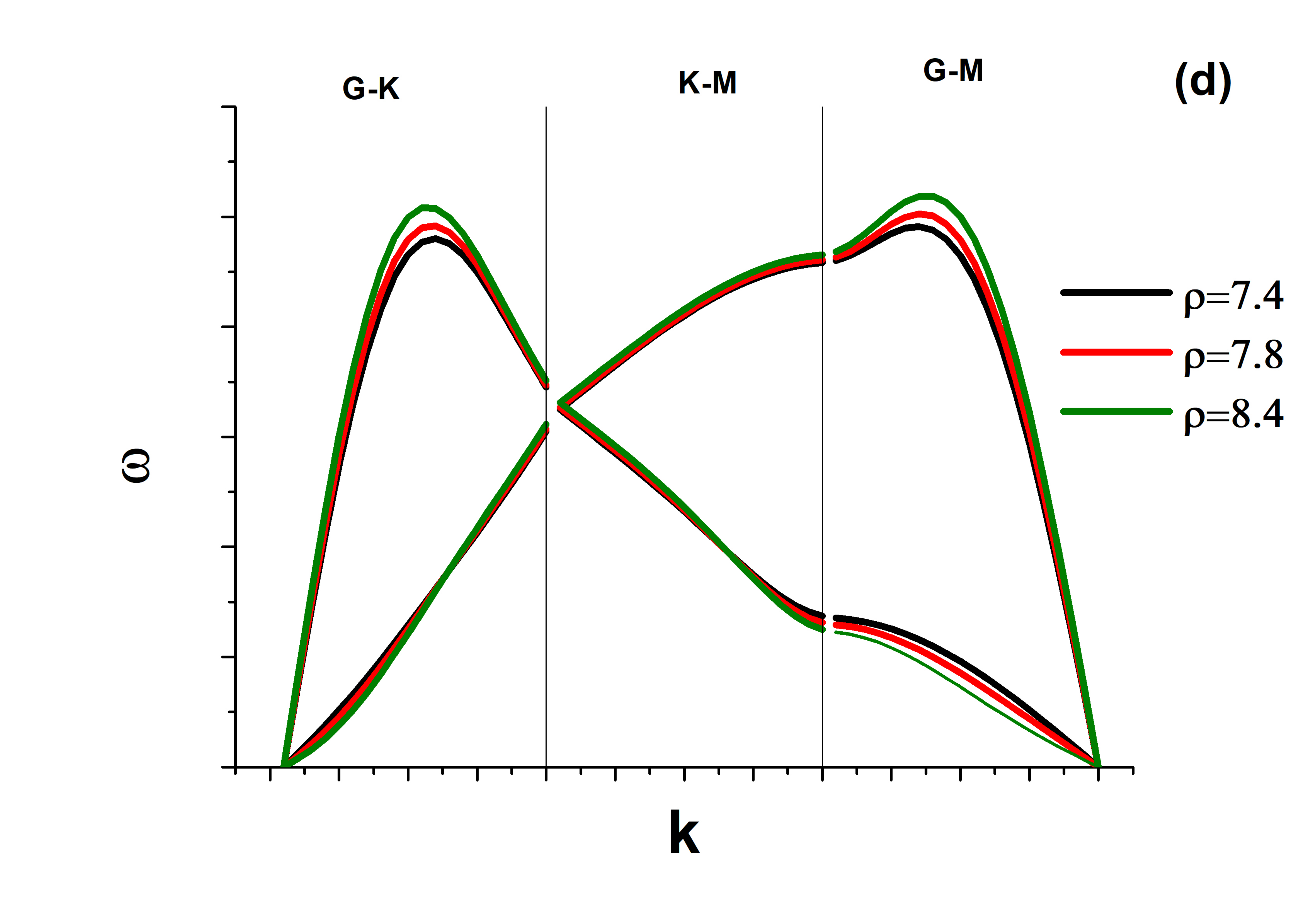}%

\caption{\label{phon} The phonon spectra of the longitudinal and
transverse modes of the Hertzian spheres with $\alpha = 7/2$ in
several density intervals.}
\end{figure}

It is known that in the case of a density anomaly in the crystal,
anomalous behavior of phonon frequencies is observed, i.e., the
frequency of phonons on one or several phonon branches can
decrease with increasing density. In a normal crystal, the
frequency of all phonon branches increases with increasing
density. We performed calculations of phonon spectra (longitudinal
and transverse modes) in several ranges of densities, namely, in
two regions where a density anomaly was detected, and in two
regions with a positive coefficient of thermal expansion. The
calculation results are shown in Figs. \ref {phon} (a) - (d). The
density areas in panels (a) and (c) correspond to normal mode. It
can be seen that in these regions the frequencies of all branches
increase with increasing density. The behavior of phonon spectra
in the areas with a density anomaly (panels (b) and (d)) is
different from normal. In both density intervals ($\rho = 3.0$ -
$3.8$ for the first area with a density anomaly and $\rho = 7.4$ -
$8.4$ for the second), the transverse modes along all directions
show negative dependence on density. This suggests a connection of
the density anomaly in the studied system with exactly the
transverse oscillations of the crystalline lattice.

\begin{figure}

\includegraphics[width=8cm]{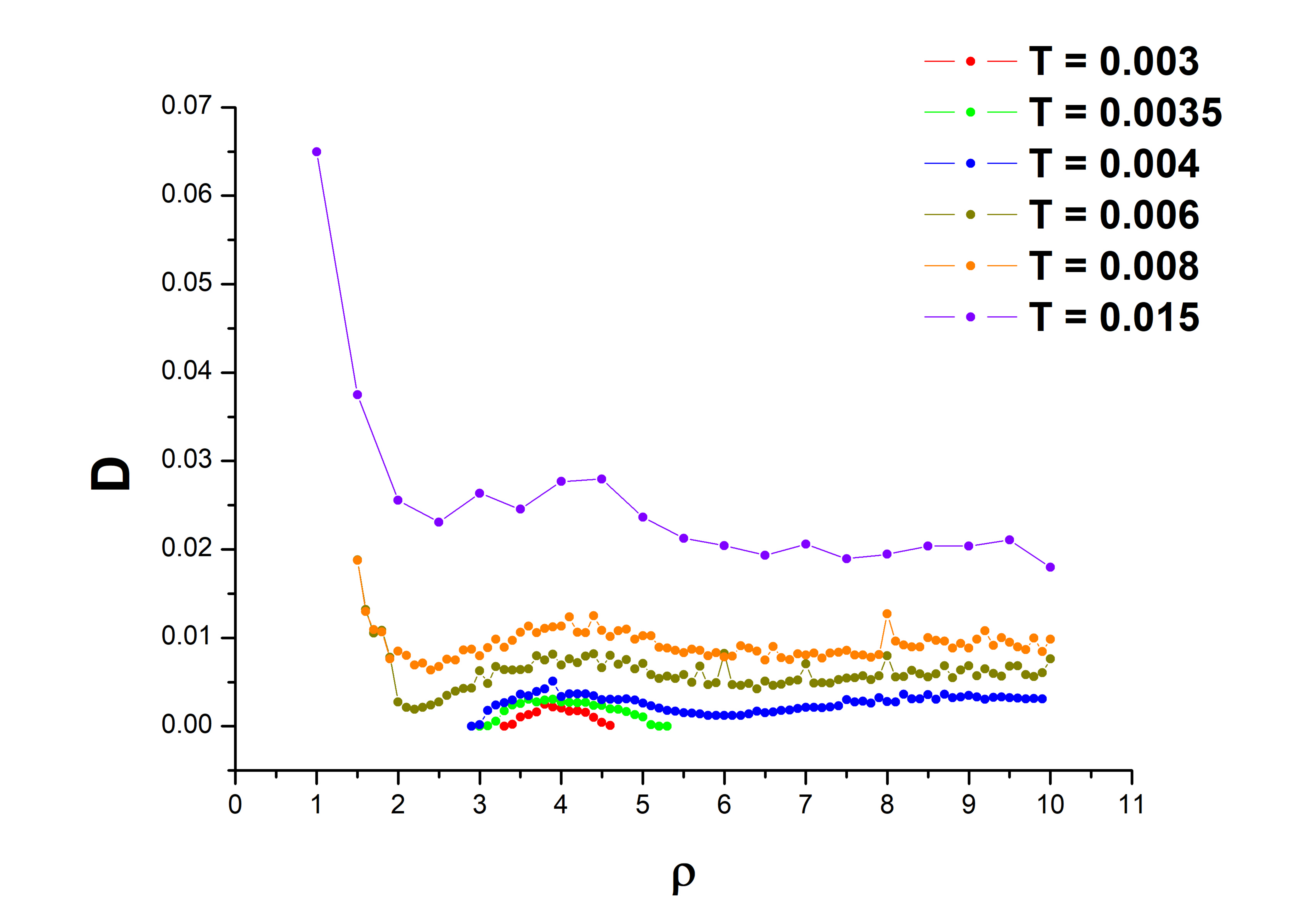}%

\caption{\label{diff} The diffusion coefficient of the system
along several isochores.}
\end{figure}

Along with the density anomaly, we found a diffusion anomaly,
i.e., an increase in the diffusion coefficient of the system with
an isothermal increase in density. Figure \ref{diff} shows the
diffusion coefficients of the system along several isotherms. It
can be seen that even at high temperatures, for example, $T =
0.015$ (the maximum melting point in the system is $T = 0.0058$)
there are two areas with an anomalous increase in the diffusion
coefficient: from density $\rho = 2.5$ to $4.5$ and from density
$6.5$ to $9.5$.

In the 2D system of Hertzian spheres with $\alpha = 5/2$, the
anomalous properties of the liquid in the reentrant melting region
\cite {molphys} between two different crystal phases can be
explained by the restructuring of the liquid as a result of
melting of the triangular crystal and crystallization of the
square one. In the case of $\alpha = 7/2$, a diffusion anomaly was
detected in the reentrant melting regions of only one triangular
phase. Let us try to explain this phenomenon using the example of
the behavior of radial distribution functions (RDF) (see Figs. 12
(b) and (c) in our paper \cite {hz72}) and translational order
parameter (TOP) $\Psi _ T$ (see Fig. 2 in \cite {hz72}) along
isotherm $T = 0.003$. From Fig. 12 (b) it can be seen that with an
increase in density to 3.7 corresponding to the anomalous
diffusion boundary at $T = 0.003$, the RDFs become less
structured, which leads exactly to an increase in the diffusion
coefficient with increasing density. With a further increase in
density and approaching the left branch of the melting line of the
second bump (see Fig. 12 (c)), the behavior of the RDFs changes to
the opposite, they become more structured, which will lead to
normal liquid behavior, that is, a decrease in the diffusion
coefficient with increasing density. According to Fig. 2, in the
anomalous diffusion region, the TOP decreases with increasing
density, which is a sign of the presence of a structural anomaly
in the system.

The areas of density and diffusion anomalies are shown on the
phase diagram in Fig. \ref {pd}. It can be seen from this figure
that the existence domain of the density anomaly is close to the
melting line, while the existence domain of the diffusion anomaly
reaches very high temperatures. This stability of the diffusion
anomaly evidences complex behavior of the dynamic properties of
the Hertzian sphere system with $\alpha = 7/2$, which was
discussed above.

\section{Conclusions}

The paper presents a molecular dynamics simulation study of
anomalous behavior of a 2D system of Hertzian spheres with
exponent $\alpha = 7/2$. Previously, it has been shown \cite
{miller, hz72} that in this system there is only one crystalline
phase, a triangular crystal, with several maxima and minima on the
melting line. We find water-like density and diffusion anomalies
in the liquid phase in the reentrant melting regions. Moreover, a
density anomaly was observed in both liquid and solid phase. We
have calculated the phonon spectra of longitudinal and transverse
modes. The transverse modes along all directions have been shown
to display negative dependence of frequency on density in the
regions with density anomalies. This indicates an association of
the density anomaly with transverse oscillations of the crystal
lattice.

The regions of density and diffusion anomalies have been drawn on
the phase diagram.

It has been found that the stability regions of anomalous
diffusion extend to temperatures well above the maximum melting
point of the triangular crystal at $T = 0.0058$.

From analysis of the translational order parameter, which
decreases with increasing density in the reentrant melting
regions, an assumption has been made that there was a structural
anomaly in the system.

\section{Acknowledgments}
This work was carried out using computing
resources of the federal collective usage centre "Complex for
simulation and data processing for mega-science facilities" at NRC
"Kurchatov Institute", http://ckp.nrcki.ru, and supercomputers at
Joint Supercomputer Center of the Russian Academy of Sciences
(JSCC RAS). The work was supported by the Russian Science
Foundation (Grant No 19-12-00092).

\end{document}